\documentclass[3p,times]{elsarticle}
\usepackage{ecrc}
\volume{00}
\firstpage{1}
\journalname{Nuclear Physics A}

\runauth{M.~Panero et~al.}
\jid{npa}
\jnltitlelogo{Nuclear Physics A}

\usepackage{amssymb}

\biboptions{square,comma,numbers,sort&compress}

\usepackage[figuresright]{rotating}

\newcommand{\SU}{\mathrm{SU}}

\newcommand{\dd}{{\rm{d}}}

\newcommand{\tr}{{\rm Tr}}

\newcommand{\qhat}{\hat{q}}
\newcommand{\qhatNLO}{\hat{q}^{\mbox{\tiny{NLO}}}}

\newcommand{\gE}{g_{\mbox{\tiny{E}}}}

\newcommand{\mE}{m_{\mbox{\tiny{E}}}}
\newcommand{\mD}{m_{\mbox{\tiny{D}}}}
\newcommand{\Cfund}{\mathcal{C}_{\mbox{\tiny{f}}}}
\newcommand{\Cadj}{\mathcal{C}_{\mbox{\tiny{a}}}}
\newcommand{\tfund}{t_{\mbox{\tiny{f}}}}

\newcommand{\eq}[1]{\begin{equation}\label{#1}}
\newcommand{\en}{\end{equation}}
\newcommand{\eqar}[1]{\begin{eqnarray}\label{#1}}
\newcommand{\enar}{\end{eqnarray}}

\begin{document}

\begin{frontmatter}

\title{Investigating jet quenching on the lattice}
\author[IFT]{Marco~Panero}
\author[HY]{Kari~Rummukainen}
\author[UR]{Andreas~Sch\"afer}

\address[IFT]{Instituto de F\'{\i}sica T\'eorica, Universidad Aut\'onoma de Madrid \& CSIC, E-28049 Cantoblanco, Madrid, Spain}
\address[HY]{Department of Physics \& Helsinki Institute of Physics, P.O. Box 64, FI-00014 University of Helsinki, Finland}
\address[UR]{Institute for Theoretical Physics,  University of Regensburg, D-93040 Regensburg, Germany}

\begin{abstract}
Due to the dynamical, real-time, nature of the phenomenon, the study of jet quenching via lattice QCD simulations is not straightforward. In this contribution, however, we show how one can extract information about the momentum broadening of a hard parton moving in the quark-gluon plasma, from lattice calculations. After discussing the basic idea (originally proposed by Caron-Huot), we present a recent study, in which we estimated the jet quenching parameter non-perturbatively, from the lattice evaluation of a particular set of gauge-invariant operators.
\end{abstract}


\end{frontmatter}

\section{Introduction}
\label{sec:intro}

Jet quenching is a very important experimental signature of the quark-gluon plasma (QGP): when a hard parton propagates in the deconfined medium, it undergoes multiple interactions with the QGP constituents, which decrease its energy and induce a transverse momentum component. Eventually, this leads to the suppression of yields at large transverse momenta and of correlations between back-to-back hadrons in particle spectra detected in heavy-ion collisions. The momentum broadening of a hard parton in the QGP can be described in terms of the phenomenological parameter $\qhat$: it represents the average increase of the squared transverse momentum component per unit length~\cite{Baier:1996kr, Baier:1996sk, Baier:2000mf}. It can be computed as the second moment of the collision kernel $C(p_\perp)$ associated with the interactions between the parton and the plasma constituents:
\begin{equation}
\label{qhat_definition}
\qhat = \frac{\langle p^2_\perp \rangle}{L} = \int \frac{ \dd^2 p_\perp}{(2\pi)^2} p^2_\perp C(p_\perp).
\end{equation}
A theoretical derivation of the jet quenching parameter $\qhat$ from first principles is very challenging, since it involves different energy scales and, consequently, non-trivial interplay of perturbative and non-perturbative dynamics. 

In principle, Monte Carlo simulations on the lattice would be an ideal tool to compute $\qhat$ numerically---especially at temperatures close to deconfinement, in which the QGP is relatively strongly coupled. Unfortunately, due to the \emph{real-time} nature of the phenomenon, a formulation of the problem on a \emph{Euclidean} lattice is far from straightforward. In this contribution, however, following an idea originally proposed by Caron-Huot~\cite{CaronHuot:2008ni}, we will discuss how it is possible to make progress in this direction, presenting the results of our recent work~\cite{Panero:2013pla}.

\section{A dimensionally reduced effective theory for high-temperature QCD}
\label{sec:dimred}

Although asymptotic freedom implies that the QCD coupling is weak for processes involving sufficiently high energy scales (like the hard thermal scale $\pi T$ characteristic of a system at high temperature $T$), a purely perturbative approach fails in thermal non-Abelian gauge theories. Even at arbitrarily high temperatures, the physics of long-wavelength modes---those at soft $O(gT)$ or ultrasoft $O(g^2T/\pi)$ scales---has non-perturbative features: this is due to the fact that infrared singularities lead to a non-trivial structure for perturbative expansions, and limit their validity to a finite order~\cite{Linde:1980ts, Gross:1980br}. At temperatures attainable in present accelerators, the contributions from non-perturbative terms are generally non-negligible. This problem can be properly addressed by means of a dimensionally reduced effective theory~\cite{Ginsparg:1980ef, Appelquist:1981vg, Nadkarni:1982kb, Kajantie:1995dw}, obtained from the formulation of equilibrium finite-temperature QCD as a four-dimensional theory (with a compact Euclidean time direction playing the r\^ole of the inverse temperature), by integrating out all non-static modes for the temporal component of the gauge field. This leads to an effective theory (electrostatic QCD, or EQCD) defined by the Lagrangian of three-dimensional $\SU(3)$ Yang-Mills theory coupled to an adjoint scalar field:
\begin{equation}
\label{continuum_EQCD}
\mathcal{L} = \frac{1}{4} F_{ij}^a F_{ij}^a + \tr \left( (D_i A_0)^2 \right) + \mE^2 \tr \left( A_0^2 \right) + \lambda_3 \left( \tr \left( A_0^2 \right) \right)^2.
\end{equation}
Fixing the dimensionful 3D gauge coupling $\gE$, the mass and quartic coupling via a \emph{matching} procedure, this effective theory describes the physics of high-temperature QCD for all modes up to the soft scale $gT$.\footnote{Integrating out the scalar field, one can obtain a further effective theory (magnetostatic QCD, or MQCD), which is just three-dimensional $\SU(3)$ Yang-Mills theory and describes the ultrasoft modes $O(g^2T/\pi)$ of QCD.}

A key observation pointed out in ref.~\cite{CaronHuot:2008ni} (see also refs.~\cite{Benzke:2012sz, Laine:2012ht, Ghiglieri:2013gia, Laine:2013lia, Cherednikov:2013pba, Nam:2014sva} for work on related ideas) is that, for a hard massless parton moving through the QGP, the contributions to $\qhat$ from collinear components of the medium fields are suppressed, and the momentum broadening would be the same even if the parton velocity exceeded the speed of light, making its trajectory space-like. This suggests that the problem can be addressed in a Euclidean setup. In fact, one can rigorously prove that the soft contribution to jet quenching can be directly computed in EQCD. Such computation was carried out perturbatively in refs.~\cite{CaronHuot:2008ni, Ghiglieri:2013gia} and non-perturbatively (through lattice simulations) in our work~\cite{Panero:2013pla}.

\section{Lattice study}
\label{sec:lattice}

Our lattice regularization of EQCD follows previous numerical studies of this theory~\cite{Hietanen:2008tv}. We run simulations for parameters corresponding to QCD with $n_f=2$ light quarks, at temperatures $T \simeq 398$~MeV and $2$~GeV. We extracted the soft contribution to momentum broadening of a hard light quark by computing two-point correlators of null Wilson lines along the $t-x_3=\emph{const}$ direction, for fixed $x_1$ and $x_2$; let $r$ be the separation between the lines (along the $x_1$ direction), and $\ell$ denote the length of the $x_3$ component of each line. The correlator is rendered gauge-invariant by including parallel transporters along the transverse direction, and tracing over the color indices of the resulting loop~\cite{Benzke:2012sz}. Upon regularization on the lattice, the null lines in this loop are mapped to ``staircase'' paths, built multiplying unitary parallel transporters $U_3(x)$ along one lattice spacing $a$ in the $x_3$ direction, and matrices $H(x)= \exp [- a \gE^2 A_0(x) ]$. The latter are Hermitian (rather than unitary) operators, which can be interpreted as the counterpart, in EQCD, of parallel transporters over paths of length $a$ in the \emph{real-time} direction. Accordingly, we arrive at the operator
\begin{equation}
\label{decorated_loop}
W(\ell, r) = \tr \left( L_3 L_1 L^{-1}_3 L^\dagger_1 \right), \qquad \mbox{with} \qquad L_3 = \prod (U_3 H), \qquad L_1 = \prod U_1
\end{equation}
(where we omitted the coordinate dependence for simplicity), which has well-defined renormalization properties~\cite{D'Onofrio:2014qxa}. We computed the expectation values of this operator using a multilevel algorithm~\cite{Luscher:2001up} and extracted the differential transverse collision kernel in coordinate space as
\begin{equation}
\label{potential}
V(r) = - \lim_{\ell \to \infty} \frac{1}{\ell} \ln\langle W(\ell, r) \rangle.
\end{equation}
Up to a trivial additive term, this quantity equals minus the transverse Fourier transform of $C(p_\perp)$ and, since (the soft contribution to) the jet quenching parameter is the second moment of the latter, it can be extracted from the curvature of $V(r)$ near the origin (see ref.~\cite{Panero:2013pla} for technical details). 

\begin{figure*}[-t]
\centerline{\includegraphics[height=0.3\textheight]{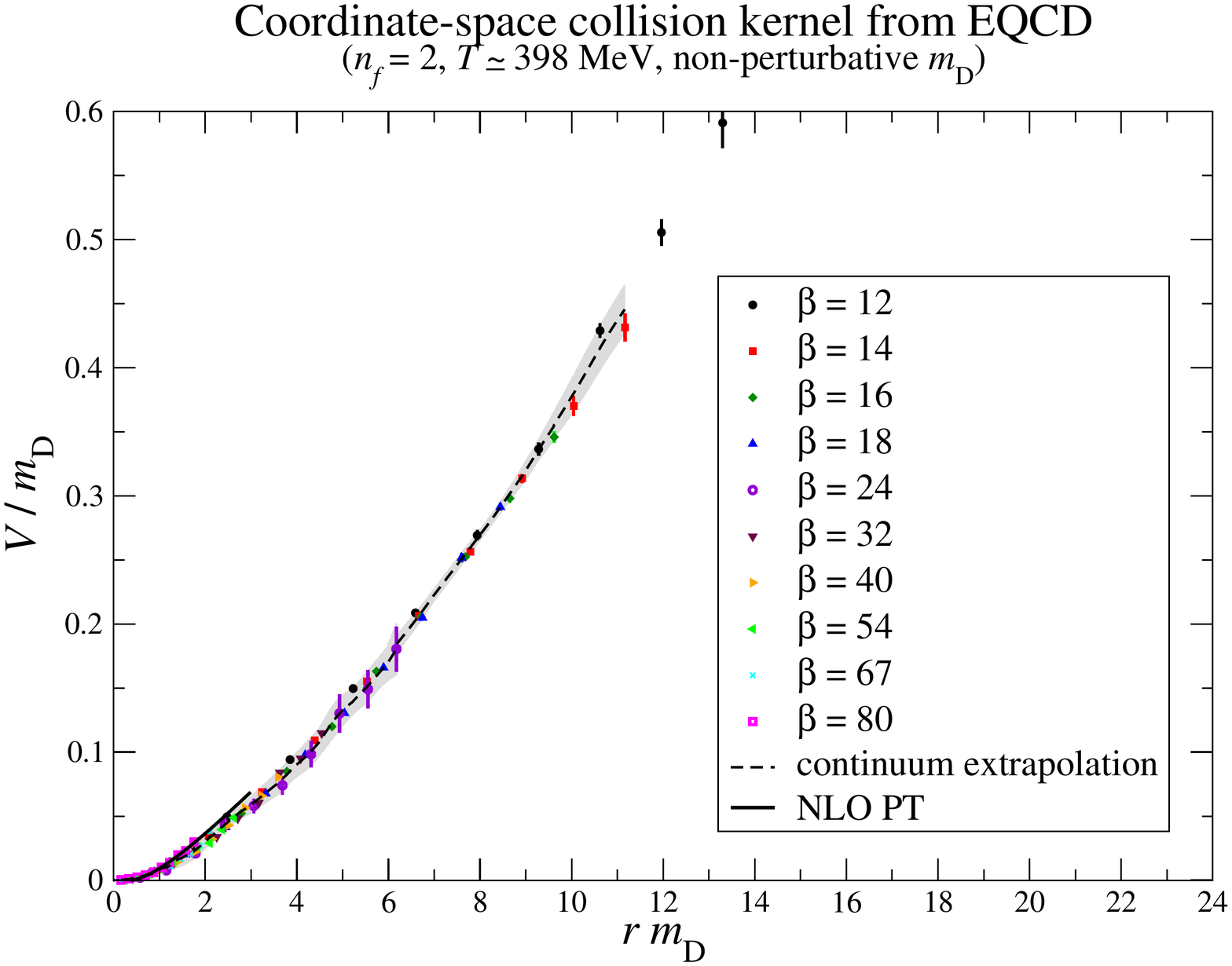} \hfill \includegraphics[height=0.3\textheight]{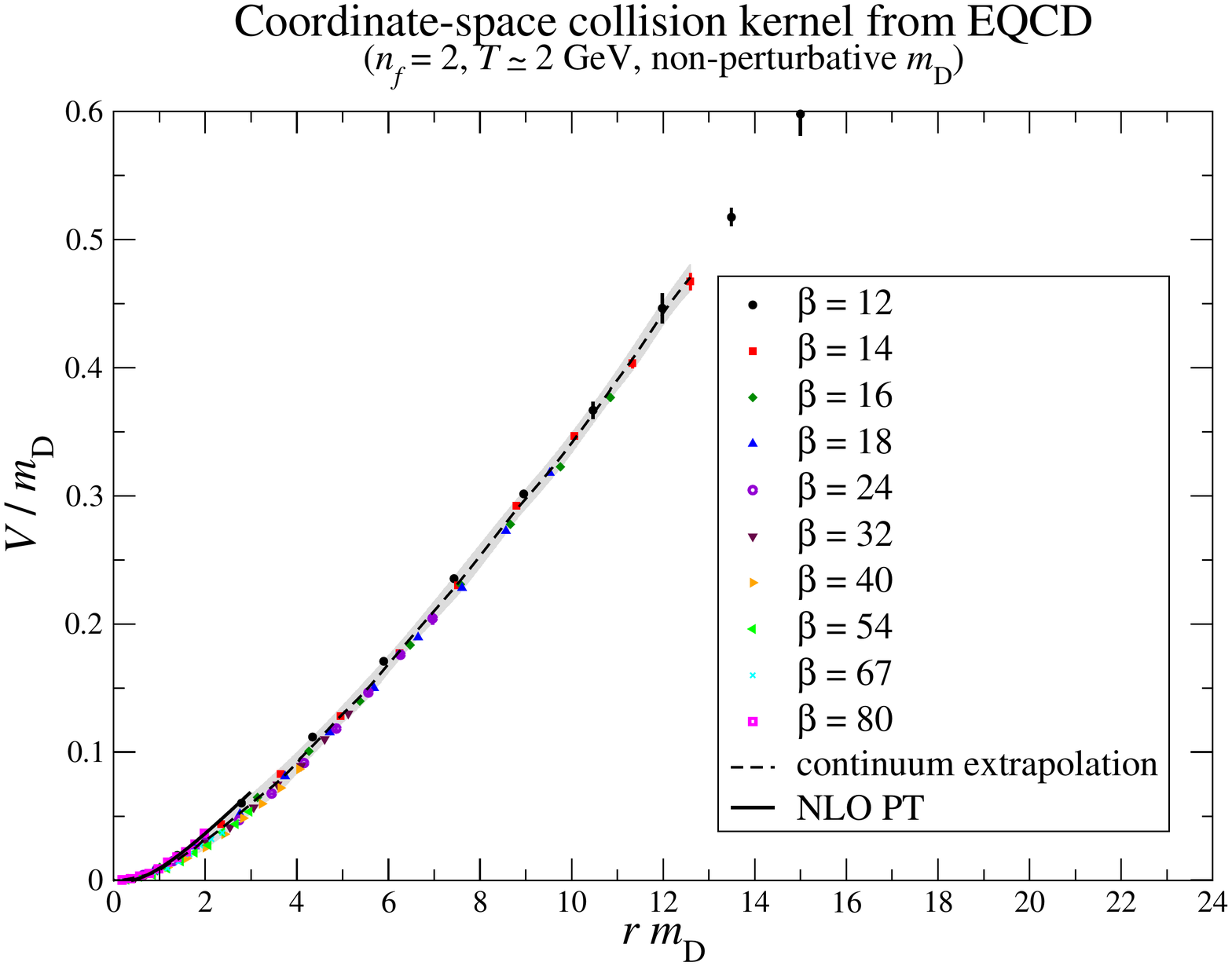}}
\caption{Transverse collision kernel $V(r)$ extracted from our EQCD simulations, at $T \simeq 398$~MeV (left-hand-side panel) and at $T \simeq 2$~GeV (right-hand-side panel). Symbols of different colors correspond to different values of $\beta = 6/(a \gE^2)$, i.e. to different values of the lattice spacing $a$. The continuum extrapolation is indicated by the dashed black line (with the gray band denoting the associated uncertainty). Both $V$ and $r$ are displayed in units of the non-perturbatively evaluated Debye screening mass $\mD$~\cite{Laine:1999hh}. The solid black curve is the NLO perturbative prediction~\cite{CaronHuot:2008ni, Ghiglieri:2013gia}.
\label{fig:V}}
\end{figure*}

The results of our simulations (in which discretization effects are under control, as data obtained from lattices of different spacing fall on a common, narrow band) show that, both at the higher and at the lower temperature that we investigated, the soft contribution to $\qhat$ is much larger than predicted perturbatively at the next-to-leading order (NLO):
\begin{equation}
\qhatNLO = g^4 T^2 \mD \Cfund \Cadj \frac{ 3 \pi^2 + 10 - 4 \ln 2 }{32 \pi^2},
\end{equation}
where $\mD = \mE + O(g^2 T)$ and $\mE = gT \sqrt{(\Cadj + n_f \tfund )/3}$, with $\Cfund=4/3$, $\Cadj=3$, and $\tfund=1/2$. This signals that contributions beyond NLO (which, due to the issues mentioned above, cannot be described by perturbation theory) are, in fact, dominant, and is hardly surprising: in the same temperature range, also the Debye mass $\mD$ receives a large contribution from non-perturbative effects~\cite{Laine:1999hh}. Interestingly, our numerical results for $V(r)$ become compatible with the analytical NLO prediction~\cite{CaronHuot:2008ni, Ghiglieri:2013gia}, if one uses the non-perturbative values for the Debye mass $\mD$: see fig.~\ref{fig:V}. This rescaling in terms of the non-perturbative Debye mass also makes the curves corresponding to $V(r)$ at the two different temperatures overlap with each other, within the uncertainties, indicating that any non-trivial temperature dependence of $V$ is inherited from $\mD$. For a physical coupling $g^2 \sim 2.6$~\cite{Laine:2005ai} at RHIC temperatures, our results lead us to estimate $\qhat \simeq 6$~GeV$^2$/fm, with an uncertainty around $15$--$20 \%$. This value is close to those obtained from holography~\cite{Liu:2006ug, Armesto:2006zv, Gubser:2006nz, CasalderreySolana:2007qw, Gubser:2008as, Hatta:2008tx, Liu:2008tz, Gursoy:2009kk, Kiritsis:2011bw, Li:2014hja} and from phenomenological models~\cite{Dainese:2004te, Eskola:2004cr} (even though the latter exhibit some dependence on the details of the underlying approach~\cite{Bass:2008rv}, and more recent studies suggest smaller values~\cite{Burke:2013yra}).

\section{Conclusions and perspectives}
\label{sec:conclusions}

To summarize, following an idea originally proposed in ref.~\cite{CaronHuot:2008ni}, in our recent work~\cite{Panero:2013pla} we showed how it is possible to extract non-perturbative information relevant for the physics of jet quenching from simulations on a Euclidean lattice. This opens up the possibility of investigating a whole class of real-time phenomena in thermal QCD from first principles, without relying on model-dependent assumptions. Interestingly, recently related ideas have also been discussed for QCD at zero temperature~\cite{Musch:2011er, Ji:2013dva, Lin_Lattice2013, Cherednikov:2014jva, Lin:2014zya}. Our approach gives direct access to the full soft contribution to the collision kernel in coordinate space $V(r)$. It is blind to contributions from hard thermal modes $O(\pi T)$ (which can be reliably estimated using perturbation theory, and are found to be subdominant), and allows a clear separation between the physics at different scales, consistently with the modern approach to finite-temperature QCD~\cite{Appelquist:1981vg, Braaten:1995cm, Braaten:1995jr, Kajantie:1995dw, Laine:2005ai}. Finally, our results are corroborated by a recent classical lattice gauge theory study~\cite{Laine:2013lia}.

The study that we presented in ref.~\cite{Panero:2013pla} could be refined using an improved formulation of the lattice action, using the multilevel algorithm recently proposed in ref.~\cite{Mykkanen:2012dv}, and/or studying the temperature dependence of $\qhat$ more in detail. In addition, it would also be interesting to check whether $\qhat$ exhibits a non-trivial dependence on the number of quark colors $N$. In particular, all holographic computations rely on the approximation of a large number of colors~\cite{'tHooft:1973jz}, which has a number of interesting properties~\cite{Lucini:2012gg, Lucini:2013qja, Panero:2012qx}. Previous lattice studies (both in four and in three spacetime dimensions)~\cite{Lucini:2005vg, Bringoltz:2005rr, Panero:2009tv, Datta:2009jn, Mykkanen:2012ri, Lucini:2012wq, Caselle:2011fy, Caselle:2011mn} already showed that the dependence on $N$ is essentially trivial for static equilibrium properties of the QGP: is this the case also for quantities related to real-time dynamics?

\noindent{\bf Acknowledgements} This work is supported by the Spanish MINECO (grant FPA2012-31686 and ``Centro de Excelencia Severo Ochoa'' programme grant SEV-2012-0249), by the Academy of Finland (project 1134018), by the German DFG (SFB/TR 55), and partly by the European Community (FP7 programme HadronPhysics3).

\vspace{-4mm}

\bibliographystyle{elsarticle-num}
\bibliography{hardprobes2013proc}

\end{document}